\documentclass[%
 reprint,%
 amssymb, amsmath,%
 prc,cha,%
superscriptaddress
]{revtex4-2}
\usepackage[colorlinks=true,linkcolor=blue]{hyperref}%
\usepackage{physics}
\usepackage{graphicx}
\usepackage{lipsum}
\usepackage{mwe}
\usepackage{caption}
\usepackage{dcolumn}% Align table columns on decimal point
\usepackage{bm}% bold math
\usepackage{adjustbox}
\usepackage{rotating}
\usepackage{xcolor}
\usepackage{ulem}
\begin{document}

\preprint{APS/123-QED}
\title{Finite temperature effects in magnetic dipole transitions}%

\author{Amandeep Kaur}\email[]{akaur.phy@pmf.hr}
\affiliation{Department of Physics, Faculty of Science, University of Zagreb, Bijeni\v{c}ka c. 32,  10000 Zagreb, Croatia}

\author{Esra Y{\"{u}}ksel}\email[]{e.yuksel@surrey.ac.uk}
\affiliation{Department of Physics, University of Surrey, Guildford, Surrey GU2 7XH, United Kingdom}

\author{Nils Paar}\email[]{npaar@phy.hr}
\affiliation{Department of Physics, Faculty of Science, University of Zagreb, Bijeni\v{c}ka c. 32,  10000 Zagreb, Croatia}

\date{\today}

\begin{abstract} 
Finite temperature effects in electromagnetic transitions in nuclei contribute to many aspects of nuclear structure and astrophysically relevant nuclear reactions. While electric dipole transitions have already been extensively studied, the temperature sensitivity of magnetic transitions remains largely unknown. This work comprises the study of isovector magnetic dipole excitations (M1) occurring between spin-orbit (SO) partner states using the recently developed self-consistent finite temperature relativistic quasiparticle random phase approximation (FT-RQRPA) in the temperature range from $T=$ 0 to 2 MeV. The M1 strength distributions of $^{40-60}$Ca and $^{100-140}$Sn isotopic chains exhibit a strong temperature dependence. The M1 strength peaks shift significantly towards the lower energies due to the decrease in SO splitting energies and weakening of the residual interaction, especially above the critical temperatures ($T_\textrm{c}$) where the pairing correlations vanish. By exploring the relevant two-quasiparticle ($2qp$) configurations contributing to the M1 strength of closed- and open-shell nuclei, new proton and neutron excitation channels between SO partners are observed in low- and high-energy regions due to the thermal unblocking effects around the Fermi level. At higher temperatures, we have noticed an interesting result in $^{40,60}$Ca nuclei, the appearance of M1 excitations, which are forbidden at zero temperature due to fully occupied (or fully vacant) spin-orbit partner states.

\end{abstract}
\maketitle
%\tableofcontents
%\section{\label{sec:level1}First-level heading:\protect\\ The line
%break was forced \lowercase{via} \textbackslash\textbackslash}

%%%%%%%%%%%%%%%%%%%%%%%%%%%%%%%%%%%%%%%%%%%%%%%%%%%%%%%%%%%%%%%%%%%%%%%%%%
\section{Introduction} \label{Intro}
A comprehensive understanding of magnetic dipole (M1) nuclear response is essential to various aspects of nuclear structure phenomena, such as isospin-mixing, isospin-splitting, and ground-state correlations \cite{Heyde82,PhysRevC.80.031302,PhysRevC.99.064329}. It also aids in the study of radiative neutron capture, which has a key role in the production of neutron-rich elements in hot stellar environments \cite{PhysRevC.94.044306,PhysRevLett.100.011101,Loens,goriely2011}. Several experimental and theoretical studies have revealed intriguing behavior in $\gamma$-ray strength functions ($\gamma$SF), and a notable enhancement is observed in the strength function toward lower transition energies \cite{PhysRevLett.89.272502,PhysRevC.20.2072,IGASHIRA1986301,PhysRevLett.52.102,PhysRev.126.671,GORIELY199810,fanto2021low,beaujeault2023zero}. The fine structure of strength functions in the $E_{\gamma}=$ 5-8 MeV region is generally determined using the low-energy part of the electric dipole (E1) response; however, some studies also suggest an anomalous increase below 4 MeV in $\gamma$SF, which is attributed to the M1 strength \cite{PhysRevC.71.044307,PhysRevLett.92.172501,PhysRevLett.93.142504,PhysRevLett.111.232504}. Thus, more research is required to explore the role of M1 excitations in the $\gamma$SF. The M1 spin-flip excitations are obtained at higher excitation energies around 8 MeV, which has so far been difficult to measure experimentally \cite{Heyde82}. Further, a so-called scissors mode of M1 excitations in deformed nuclei is observed at energies around 3 MeV due to the scissors-like motion of neutrons and protons relative to each other \cite{Bohle01,Bohle02,Pietralla1998,PhysRevLett.92.172501}.

M1 excitations exhibit a wide range of characteristics related to their energies, transition strengths, and decay properties, which are expected to be highly sensitive to the extreme conditions of temperature ($T$), isospin (N/Z), and deformation ($\beta$) of nuclei. Recently, the relativistic quasiparticle random phase approximation (RQRPA) has been utilized to investigate the role of residual interaction, spin-orbit (SO) splitting, pairing correlations, and neutron excess on M1 response \cite{Oishi105,Oishi_20,PhysRevC.102.044315,PhysRevC.103.054306,oishi57}. 
However, the knowledge about thermal effects on M1 excitations is rather limited. Temperature can significantly impact the electromagnetic nuclear response, which in turn modifies astrophysically relevant quantities such as neutron capture cross-section, nuclear reaction rates, and element abundances.
 
In previous studies, temperature effects in electric multipole excitations have been investigated using several extensions of the random phase approximation (RPA) \cite{KHAN200431,PhysRevC.96.024303,yuksel2019nuclea,NIU200931,PhysRevC.88.031302}. 
The finite temperature quasiparticle RPA based on the Skyrme functional was used in studies of the electric multipole responses of hot nuclei in Refs.  
\cite{KHAN200431,PhysRevC.96.024303,yuksel2019nuclea}. A self-consistent finite temperature relativistic RPA (FT-RRPA), based on meson-exchange interaction, was successfully employed to study the evolution of isoscalar and isovector electric multipole modes with temperature; however, the pairing correlations were not implemented in this approach \cite{NIU200931}. Furthermore, finite-temperature relativistic time-blocking approximation (FT-RTBA) approach, based on Matsubara Green's function formalism, has been developed to investigate the electric nuclear response in excited nuclei \cite{PhysRevLett.121.082501,litvinova2019nuclea}. 
Lately, electromagnetic strength distributions in both closed-shell and open-shell nuclei have been obtained from first principles at zero and finite temperatures, taking into account pairing and deformation effects \cite{beaujeault2023zero}. Additionally, a fully self-consistent finite-temperature relativistic quasiparticle RPA (FT-RQRPA) has been introduced for non-charge exchange excitations at finite temperatures, which also includes pairing correlations to address open-shell nuclei \cite{kaur2023electric}. It has been employed to study the thermal effects on isovector E1 excitations, demonstrating how the isovector giant dipole resonance and low-energy dipole excitations evolve with increasing temperature. It has been shown that new low-energy states appear due to the thermal unblocking effects. Consequently, the FT-RQRPA provides new perspectives for the microscopic calculation of $\gamma$SF at finite temperatures associated with nuclear phenomena in stellar environments. Therefore, it is of particular interest to analyze the thermal effects on M1 excitations as well as their contributions to $\gamma$SF relevant for nucleosynthesis. In this work, we focus on M1 excitations, characterized by unnatural parity transitions, for which additional modifications of the residual FT-RQRPA interaction are required. More details about the RQRPA formalism for magnetic transitions and their properties in the zero-temperature limit can be found in Refs. \cite{PhysRevC.102.044315, Oishi105, Oishi_20, PhysRevC.103.054306, oishi57, kruvzic2023magneti}.

The novelty of the present study lies in its exploration of the temperature dependence of isovector (IV) M1 ($J^\pi=1^+$) excitations using the newly developed FT-RQRPA framework for nuclei in the $^{40-60}$Ca and $^{100-140}$Sn isotopic chains. To study the unnatural parity excitations of M1 type, the FT-RQRPA residual interaction is further extended by introducing the relativistic isovector-pseudovector (IV-PV) contact interaction \cite{PhysRevC.102.044315}. The primary goals of this work are (i) to investigate the evolution of M1 strength distributions with increasing temperature and isospin of nuclei; (ii) to study the impact of temperature on the SO splitting energies and their relation with the M1 nuclear response; (iii) to examine the isovector M1 non-energy-weighted and energy-weighted summations as a function of temperature and mass number; and (iv) to identify the new M1 proton and neutron excitations in the low-energy as well as the high-energy region. 

The paper is organised as follows: a brief description of the FT-RQRPA framework used in this study is given in Sec. \ref{methodology}. The extension of the residual interaction for unnatural parity M1 excitations is explained. 
Sec. \ref{results} presents the FT-RQRPA results and discussions of M1 excitations in Ca and Sn isotopic chains in the temperature range $T=0-2$ MeV. Finally, a summary of the results and conclusions are outlined in Sec. \ref{Summary}.

%%%%%%%%%%%%%%%%%%%%%%%%%%%%%%%%%%%%%%%%%%%%%%%%%%%%%%%%%%%%%%%%%%%%%%%%%%%%

\section{Methodology} \label{methodology}
In this work, we extend the FT-RQRPA from Ref. \cite{kaur2023electric} for the study of unnatural parity excitation of M1 type. 
The nuclear properties are described within the finite temperature Hartree-Bardeen-Cooper-Schrieffer (FT-HBCS) framework \cite{GOODMAN19813,yuksel2014effct}. In both the FT-HBCS and FT-RQRPA, the relativistic energy density functional (REDF) with point coupling DD-PCX interaction is implemented \cite{PhysRevC.99.034318}. The point-coupling REDF is determined from the Lagrangian density,  
\begin{equation}\label{Lagrangian}
\mathcal{L}=\mathcal{L}_{\textrm{PC}}+\mathcal{L}_{\textrm{IV-PV}},   
\end{equation}
where $\mathcal{L}_{\textrm{PC}}$ includes fermion contact interaction terms as isoscalar-scalar, isoscalar-vector and isovector-vector channels, for the detailed information see Refs. \cite{PhysRevC.99.034318,NIKSIC20141808}. The Lagrangian density (\ref{Lagrangian}) also includes the relativistic isovector-pseudovector (IV-PV) contact interaction, which is necessary for the FT-RQRPA residual interaction for the unnatural parity excitations of M1 type \cite{PhysRevC.102.044315},
\begin {equation}
\begin{aligned}
\mathcal{L}_{\textrm{IV-PV}} = -\frac{1}{2}\alpha_{\textrm{IV-PV}} \lbrack  \bar{\Psi}_{N} \gamma^{5} \gamma^{\mu} \vec{\tau}  \Psi_{N} \rbrack \cdot \lbrack  \bar{\Psi}_{N} \gamma^{5}  \gamma_{\mu} \vec{\tau} \Psi_{N}  \rbrack.
\end{aligned}
\end {equation}
The coupling strength parameter $\alpha_{\textrm{IV-PV}}=0.63$ MeVfm$^3$ for DD-PCX \cite{PhysRevC.99.034318} parameterization is obtained by minimizing the relative error $\Delta$ $\lesssim$ 1 MeV between experimentally determined M1 peak position and theoretically calculated centroid energies for magic nuclei $^{48}$Ca and $^{208}$Pb \cite{PhysRevC.102.044315,kruvzic2023magneti}. Note that the pseudovector type of interaction has been modeled as a scalar product of two pseudovectors, which leads to the parity-violating mean field at the Hartree level. Thus, it does not make a contribution to the solution of natural-parity states, including the 0$^+$ nuclear ground state. However, $\mathcal{L}_{\textrm{IV-PV}}$ has a finite contribution in the FT-RQRPA residual interactions for M1 excitations.

At finite temperature, the occupation probabilities of single-particle states are given as 
\begin{equation}
n_i=v_{i}^{2}(1-f_{i})+u_{i}^{2}f_{i},   
\end{equation}
where $u_i$ and $v_i$ are the BCS amplitudes. The temperature-dependent Fermi-Dirac distribution function is given by 
\begin{equation}
f_{i}=[1+exp(E_{i}/k_{\textrm{B}}T)]^{-1},   
\end{equation}
$k_{\textrm{B}}$ is the Boltzmann constant and $T$ is the temperature. The quasiparticle energy of a state is given by
\begin{equation}
E_i=\sqrt{(\varepsilon_i-\lambda_q)^2+\Delta_i^2},    
\end{equation}
where $\varepsilon_i$ denotes the single-particle energies and $\lambda_q$ represents the chemical potentials for either proton or neutron states. $\Delta_i$ refers to the pairing gap of the given state. A separable form of pairing interaction is introduced both in the FT-HBCS and FT-RQRPA, also including the same relativistic point coupling interaction, DD-PCX \cite{PhysRevC.80.024313}. The values of critical temperatures ($T_\textrm{c}$), where the pairing correlations vanish, are calculated using the FT-HBCS with DD-PCX interaction. For instance, the $T_\textrm{c}$ values of open-shell $^{44}$Ca, $^{52}$Ca, $^{56}$Ca, $^{64}$Ca, $^{108}$Sn, $^{116}$Sn, $^{124}$Sn, and $^{140}$Sn nuclei are obtained as 0.862, 0.528, 0.743, 0.700, 0.872, 0.834, 0.764, and 0.644 MeV, respectively. 

The non-charge exchange FT-RQRPA matrix is given by
\begin{equation}
\left( { \begin{array}{cccc}\label{eq:qrpa}
 \widetilde{C} & \widetilde{a} & \widetilde{b} & \widetilde{D} \\
 \widetilde{a}^{+} & \widetilde{A} & \widetilde{B} & \widetilde{b}^{T} \\
-\widetilde{b}^{+} & -\widetilde{B}^{\ast} & -\widetilde{A}^{\ast}& -\widetilde{a}^{T}\\
-\widetilde{D}^{\ast} & -\widetilde{b}^{\ast} & -\widetilde{a}^{\ast} & -\widetilde{C}^{\ast}
 \end{array} } \right)
 \left( {\begin{array}{cc}
\widetilde{P}  \\
\widetilde{X }  \\
\widetilde{Y}  \\
\widetilde{Q} 
 \end{array} } \right)
 = E_{w}
  \left( {\begin{array}{cc}
\widetilde{P}  \\
\widetilde{X}  \\
\widetilde{Y}  \\
\widetilde{Q} 
\end{array} } \right), \end{equation}
where $ E_{w}$ denotes the excitation energies, and $\widetilde{P}, \widetilde{X}, \widetilde{Y}, \widetilde{Q}$ depict the eigenvectors. The detailed description of the finite temperature QRPA matrices is given in \cite{SOMMERMANN198316,PhysRevC.96.024303,yuksel2019nuclea}. The FT-RQRPA matrices are diagonalized in a self-consistent way, allowing for a detailed analysis of each excitation on a state-by-state basis. We note that the $A$ and $B$ matrices contribute at $T=$ 0 and at $T \neq$ 0 as well; however, other matrix elements begin to contribute only at finite temperature due to the changes in occupation factors as well as the temperature factors. The reduced transition probability is given by

\begin{widetext}
\begin{equation}
\begin{split}
B(\textrm{M}J)=\bigl|\langle w ||\hat{F}_{J}||\widetilde0\rangle \bigr|^{2}
&=\biggl|\sum_{c\geq d}\Big\{(\widetilde{X}_{cd}^{w} + (-1)^{j_{c}-j_{d}+J}\widetilde{Y}_{cd}^{w})(u_{c}v_{d}+(-1)^{J}v_{c}u_{d})\sqrt{1-f_{c}-f_{d}} \\
&+(\widetilde{P}_{cd}^{w}+(-1)^{j_{c}-j_{d}+J}\widetilde{Q}_{cd}^{w})(u_{c}u_{d}-(-1)^{J}v_{c}v_{d})\sqrt{f_{d}-f_{c}}\Big\}\langle c ||\hat{F}_{J}||d\rangle\biggr|^{2},
\end{split}
\label{bel}
\end{equation}
\end{widetext}
where $|w\rangle$ is the excited state and $|\widetilde0\rangle$ is the correlated FT-RQRPA vacuum state, and $\hat{F}_{J}$ is the transition operator of the relevant excitation. To evaluate the magnetic dipole strength, we use the IV-M1 operator for $k$th nucleon as
\begin{equation}\label{M1 operator}
\hat{\mu}_{w}^{M1,IV}=\mu_{N}\sqrt{\frac{3}{4\pi}}\sum_{k\in A}[g_s^{IV}\hat{s}_{w}(k)+g_l^{IV}\hat{l}_{w}(k)]\hat{\tau}_0(k),
\end{equation}
including the spin $\hat{s}_{w}$ and orbital angular
momentum $\hat{l}_{w}$. The isospin convention is used as $\hat{\tau}_0(k)$ = 1 (-1) for protons (neutrons). $\mu_N=e\hbar/(2m_N)$ denotes the nuclear magneton, and the nuclear spin and orbital $g$ factors for the IV-M1 mode are $g_s$ = 4.706 and $g_l$ = 0.5 (see Refs. \cite{PhysRevC.102.044315,PhysRevC.67.034312} for more details). 
The total M1 transition strength can also be written in a compact form to identify the role of particular proton or neutron configurations,
\begin{equation}\label{Part.Contri}
B(\textrm{M1}, E_w)=\biggl|\sum_{c\geq d} \big[ b^{\pi}_{cd}(E_w)+b^{\nu}_{cd}(E_w) \big]\biggr|^2.
\end{equation}
Here, $E_w$ is excitation energy obtained from the FT-RQRPA discrete M1 spectra. The $b^{\pi}_{cd}$($E_w$) and $b^{\nu}_{cd}(E_w)$ represent the proton ($\pi$) and neutron ($\nu$) partial contributions for a specific configuration. The two-quasiparticle cut-off energy for the configuration space $E_{cut}$ is selected as 100 MeV, to provide a sufficient convergence in the M1 excitation strength. For the presentation of the results, the discrete FT-RQRPA spectrum of M1 response is smoothed with a Lorentzian function of $\Gamma=1.0$ MeV width \cite{PhysRevC.67.034312}. 

%%%%%%%%%%%%%%%%%%%%%%%%%%%%%%%%%%%%%%%%%%%%%%%%%%%%%%%%
\section{Results} \label{results}
\subsection{Calcium isotopes}
\begin{figure}[!ht]
\begin{center}
\includegraphics[width=\linewidth,clip=true]{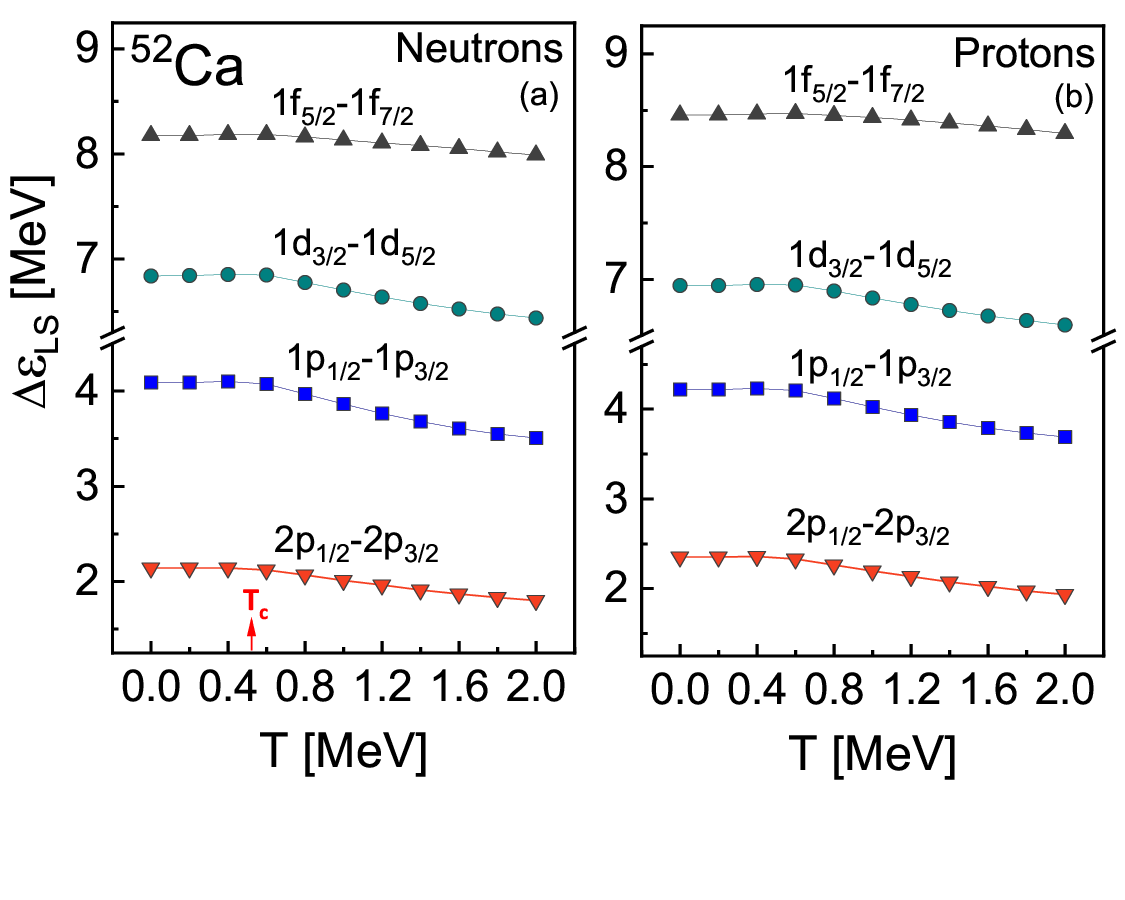}
  \end{center} \vspace{-1.5cm}
  \caption{The spin-orbit splitting energies $\Delta \varepsilon_{LS}$ of neutron (a) and proton (b) configurations as a function of temperature for $^{52}$Ca nucleus.}\label{fig1}
\end{figure}

The M1 excitation at the leading one-body operator level would take place between the spin-orbit (SO) partner orbits, provided the independent single-particle picture is a good approximation \cite{Oishi_20}. Thus, the M1 response provides important information on the underlying SO splittings. However, in the REDF framework, the M1 properties depend not only on the SO splittings but also on the effects of the residual interaction in the RQRPA \cite{Oishi_20}. 
The SO splitting energies are calculated as 
\begin{equation}
\label{gap}
\Delta \varepsilon_{LS}=\varepsilon_{nlj_{<}}-\varepsilon_{nlj_{>}},
\end{equation}
where ($nlj$) are the quantum numbers of major SO-partner single-particle states, and $j_<=l-1/2$ and $j_>=l+1/2$. In Figs. \ref{fig1}(a) and \ref{fig1}(b), the SO splitting energies are displayed both for neutrons and protons as a function of temperature for $^{52}$Ca. We note that the single-particle energies can easily be affected by the inclusion of temperature in the calculations, which, in turn, modifies the SO splitting. It is clearly seen from Fig. \ref{fig1} that the SO splitting energy is temperature-dependent. While the SO splitting energy remains almost constant up to the critical temperature $T_\textrm{c}$ = 0.528 MeV, it gradually starts decreasing at higher temperatures. This reduction of the SO splitting energies above $T_\textrm{c}$ will modify the M1 response in both the low-energy and high-energy regions, as we will discuss below.

%%%%%%%%%%%%%%%%%%%%%%%%%%%%%%%%%%%%%%%%%%%%%%%%%%%%%%%%%%%%%%
%%%%%%%%%%%%%%%%%%%%%%%%%%%%%%%%%%%%%%%%%%%%%%%%%%%%%%%%%%%%%

\begin{figure}[!ht]
  \begin{center}
\includegraphics[width=\linewidth,clip=true]{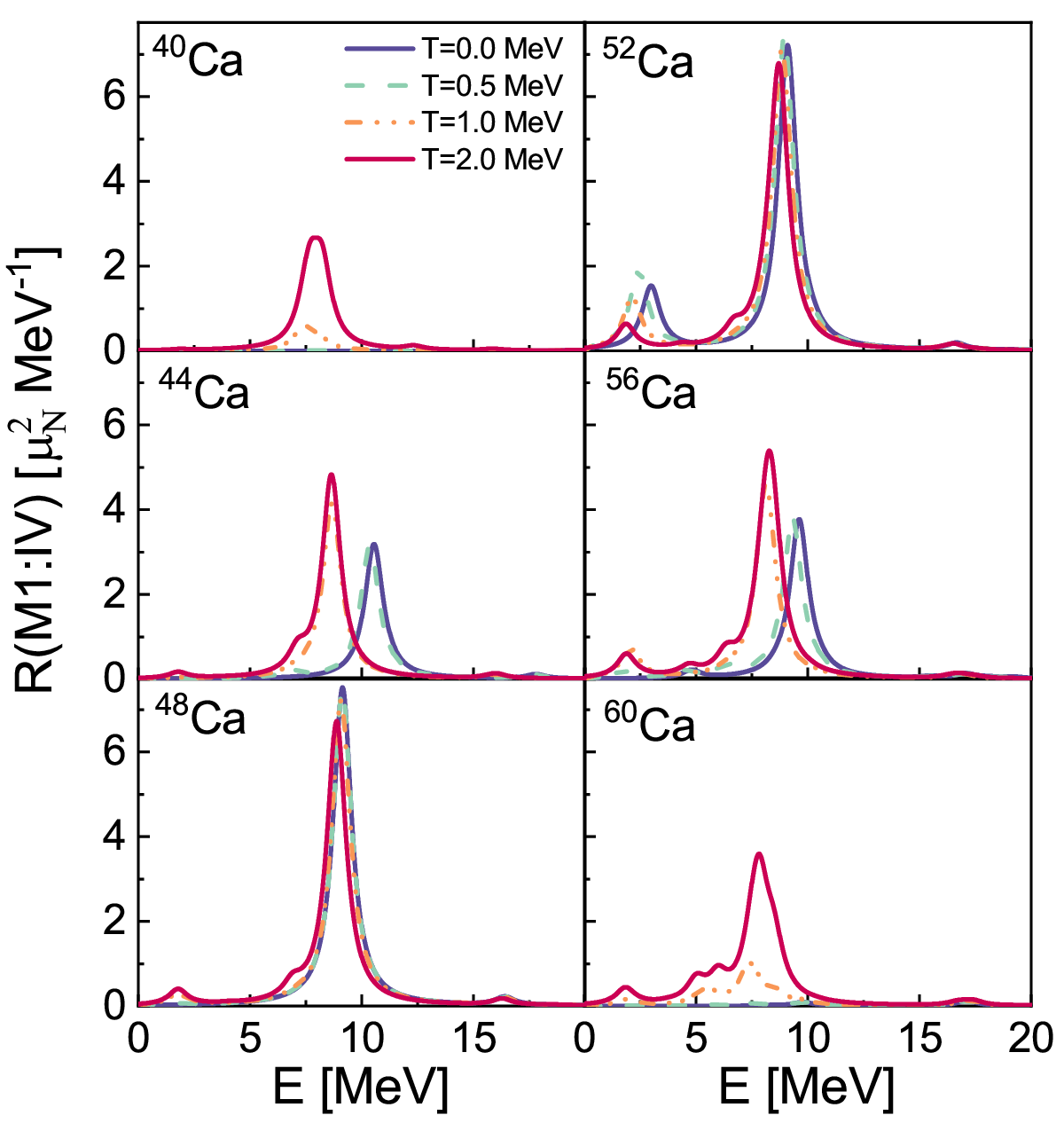}
  \end{center} \vspace{-0.5cm}
  \caption{The isovector M1 transition strength distributions of $^{40-60}$Ca isotopes, calculated using the FT-RQRPA at temperatures $T=0$, $0.5$, $1$, and $2$ MeV.}
  \label{fig2}
\end{figure}

In Fig. \ref{fig2}, we display the M1-transition strength distributions of $^{40-60}$Ca nuclei with increasing temperature. First, we consider the results at $T=0$ MeV for $^{40}$Ca $(Z,N=20)$, showing the absence of any M1-response, as ($1p_{3/2}$, $1p_{1/2}$) and ($1d_{5/2}$, $1d_{3/2}$) states are fully occupied for protons and neutrons; therefore, no SO partners are available for M1 transition. Similarly, for $^{60}$Ca with $N=40$ neutrons, states up to $1f_{5/2}$ and $2p_{1/2}$ are fully occupied, and hence the M1 transition is forbidden. Thus, no M1-response is obtained for these two nuclei at zero temperature. For $^{44-56}$Ca, one can see a strong peak in each isotope that attributes to the M1 excitation of valence neutron transitions $\nu$($1f_{7/2}$ $\rightarrow$ $1f_{5/2}$), whereas M1 transitions are not present for protons due to the shell closure at $Z=20$. A low-energy M1 peak is also observed in $^{52,56}$Ca nuclei due to the $\nu$($2p_{3/2}$ $\rightarrow$ $2p_{1/2}$) transition. More details about the evolution of M1 strength along the $^{40-60}$Ca isotopic chain at zero temperature are given in Ref. \cite{Oishi_20}.  
%%%%%%%%%%%%%%%%%%%%%%%%%%%%%%%%%%%%%%%%%%%%%%%%%%%%%%%%%%%%%%%%%%%%%%%%%%%%
%%%%%%%%%%%%%%%%%%%%%%%%%%%%%%%%%%%%%%%%%%%%%%%%%%%%%%%%%%%%%%%%%%%%%%%%%%%
%%%%%%%%%%%%%%%%%%%%%%%%%%%% M1 excitation Tables 52Ca %%%%%%%%%%%%%%%%%%%%
{\begin{table*}[ht] \renewcommand{\arraystretch}{1.2}
\tabcolsep 0.2cm
\caption{The isovector partial contributions $b_{2qp}^{\pi(\nu)}$ [$\mu_N$] of protons ($\pi$) and neutrons ($\nu$) to the dominant high-energy M1 excitations in $^{52}$Ca nucleus at $T=0$, 1 and 2 MeV are listed at corresponding excitation energies ($E$). The total $B(M1,E)$ values are obtained using Eq. (\ref{Part.Contri}), by summing over all proton and neutron configurations, also including those not listed in the table. The major contributions are highlighted in bold.}
\centering 
\begin{tabular}{l c c c c c c c} 
\hline \\[-1.0em]
   %\multicolumn{10}{c}{$b_{ph}^{\pi(\nu)}$ in \emph{high-lying} region}  \\  \hline
      Configuration                       & $T=0$ MeV    & \multicolumn{2}{c}{$T=1$ MeV}       &       & \multicolumn{3}{c}{$T=2$ MeV}  \\  
                                                         \cline{3-4}  \cline{6-8} 
                                    &$E=$ 9.09 MeV  &$E=$ 8.84 MeV &$E=$ 6.85  MeV &&$E=$ 8.71 MeV &$E=$ 8.24  MeV &$E=$ 6.68 MeV  \\ %[1ex]
\hline \\[-1.0em]
$\nu$(1f$_{7/2}$$\rightarrow$1f$_{5/2}$) &\bf{-3.363} &\bf{-3.207} &\bf{0.216} & &\bf{-2.397} &\bf{-1.694}&\bf{0.382}             \\
$\nu$(2p$_{3/2}$$\rightarrow$2p$_{1/2}$) &  -0.029    & -0.012     & -0.003    & & -0.005     &           & -0.003              \\
$\nu$(1f$_{5/2}$$\rightarrow$2f$_{7/2}$) &            & -0.006     &           & & 0.012      &0.021      & -0.003              \\
$\nu$(1g$_{9/2}$$\rightarrow$2g$_{7/2}$) &            & -0.002     &           & & -0.082     & 0.039     & 0.004               \\
$\pi$(1f$_{7/2}$$\rightarrow$1f$_{5/2}$) &            &\bf{-0.117} & 0.005     & &\bf{-0.615} &\bf{1.009} & 0.060              \\
$\pi$(1d$_{5/2}$$\rightarrow$1d$_{3/2}$) &            & -0.027     &\bf{-0.557}& &\bf{-0.115} & -0.028    &\bf{-1.189}         \\ \hline
total $B(M1,E)$[$\mu_{N}^{2}$]           & 11.262     & 11.032     & 0.115     & &  10.045    & 0.416     & 0.565               \\
                 
\hline \\ [-1.ex]
\end{tabular}
\label{table1}
\end{table*}
%%%%%%%%%%%%%%%%%%%%%%%%%%%%%%%%%%%%%%%%%%%%%%%%%%%%%%%%%%%%%%%%%%%%%%%%%%%%%

\begin{table*}[ht] \renewcommand{\arraystretch}{1.2}
\tabcolsep 0.1cm
\caption{Same as Table \ref{table1}, but for the low-energy M1 excitations in the $^{52}$Ca nucleus for $E < 5$ MeV.}
\centering 
\begin{tabular}{l c c c c } 
\hline \\[-1.0em]
  % \multicolumn{5}{c}{$b_{2qp}^{\pi(\nu)}$ in \emph{low-lying} region}  \\ \hline
   Configuration                         & $T=0$ MeV     & $T=1$ MeV   & \multicolumn{2}{c}{$T=2$ MeV}  \\
                                             \cline{4-5}
                                         & $E=$ 2.98 MeV  & $E=$ 2.14 MeV & $E=$ 4.36 MeV    & $E$=1.86 MeV\\ \hline
$\nu$(2p$_{3/2}$$\rightarrow$2p$_{1/2}$) &\bf{-1.726}    &\bf{1.520}    & -0.002       & \bf{1.057} \\
$\nu$(1f$_{7/2}$$\rightarrow$1f$_{5/2}$) &\bf{0.192}     &\bf{-0.123}   & 0.028        & -0.067 \\
$\nu$(1g$_{9/2}$$\rightarrow$1g$_{7/2}$) &               &              & \bf{-0.364}  & -0.002 \\
$\pi$(1d$_{5/2}$$\rightarrow$1d$_{3/2}$) &               & -0.006       & 0.005        & -0.018 \\
$\pi$(1f$_{7/2}$$\rightarrow$1f$_{5/2}$) &               & -0.004       & 0.004        & -0.013\\ 
$\pi$(2p$_{3/2}$$\rightarrow$2p$_{1/2}$) &               &              &              & -0.012 \\ \hline
total $B(M1,E)$ [$\mu_{N}^{2}$]          & 2.336         & 1.923        & 0.108        & 0.892 \\
    
\hline \\ [-1.ex]
\end{tabular}
\label{table2} 
\end{table*}
%%%%%%%%%%%%%%%%%%%%%%%%%%%%%%%%%%%%%%%%%%%%%%%%%%%%%%%%%%%%%%%%
%%%%%%%%%%%%%%%%%%%%%%%%%%%%%%%%%%%%%%%%%%%%%%%%%%%%%%%%%%%%%%%%
\begin{figure}[!ht]
  \begin{center}
\includegraphics[width=\linewidth,clip=true]{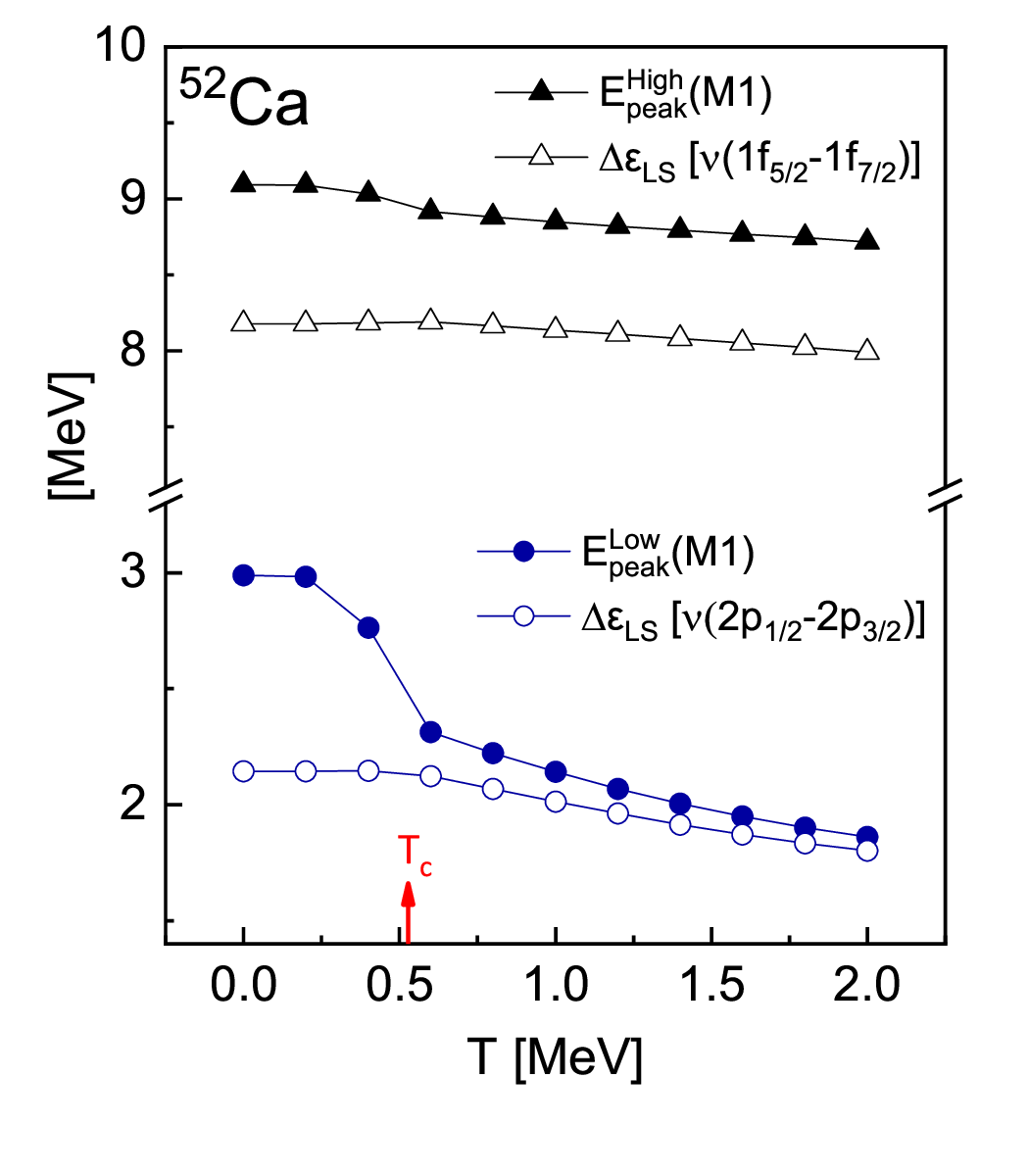}
  \end{center} \vspace{-1.0cm}
  \caption{Major high-energy ($E_{\text{peak}}^{\text{High}}$) and low-energy ($E_{\text{peak}}^{\text{Low}}$) M1 peak excitation energies, along with the SO splitting energies ($\Delta \varepsilon_{LS}$) of the respective primary contributing neutron configurations in $^{52}$Ca. The results are shown as a function of temperature in the range from $T=0$ to $2$ MeV.}
  \label{fig3}
\end{figure}
%%%%%%%%%%%%%%%%%%%%%%%%%%%%%%%%%%%%%%%%%%%%%%%%%%%%%%%%%%%%%%%
%%%%%%%%%%%%%%%%%%%%%%%%%%%%%%%%%%%%%%%%%%%%%%%%%%%%%%%%%%%%%%%
\begin{figure}[ht!]
\includegraphics[width=\linewidth,clip=true]{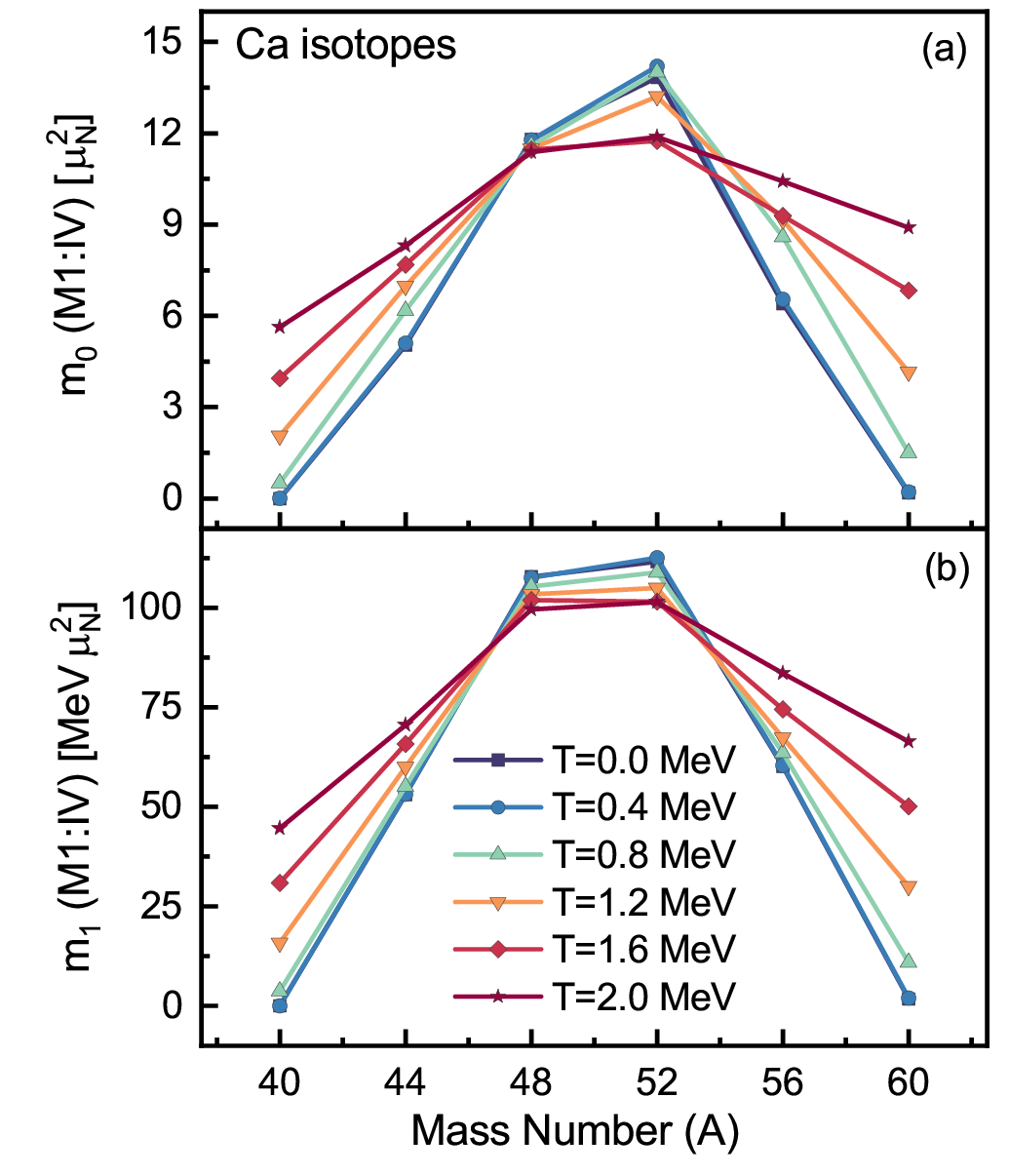}
\vspace{-0.5cm}
  \caption{(a) Non-energy-weighted $m_0$ and (b) energy-weighted $m_1$ moments of M1 response in $^{40-60}$Ca between $T=0-2$ MeV.}
  \label{fig4}
\end{figure}
%%%%%%%%%%%%%%%%%%%%%%%%%%%%%%%%%%%%%%%%%%%%%%%%%%%%%%%%%%%%%%%%

Further, it is clearly evident from Fig. \ref{fig2} that M1 transition strength distributions are remarkably sensitive to changes in temperature. At $T$ = 0.5 MeV, the results do not change much in Ca isotopes. By increasing temperature further, at $T$ = 1 and 2 MeV, an interesting outcome is observed for the case of $^{40}$Ca and $^{60}$Ca. The M1 transition strength suddenly appears for these isotopes due to the emergence of new transitions in the $\nu,\pi$ ($1d_{5/2}$, $1d_{3/2}$) and $\nu,\pi$ ($1f_{7/2}$, $1f_{5/2}$) configurations. This occurs because particles are promoted to higher-energy orbits as a result of temperature effects, leading to the thermal unblocking of forbidden M1 transitions. It is also noticed in Fig. \ref{fig2} that M1 response shifts up to $\approx$ 2 MeV to the lower energies for $^{44,48,52,56}$Ca with increasing temperature. In addition to the decrease in the pairing correlations for open-shell nuclei and softening in the repulsive residual interaction, the changes in the M1 response are also linked to the SO splitting energies, which reduce with increasing temperature, as shown in Fig. \ref{fig1}. Due to the thermal unblocking of states, new proton and neutron transitions also become possible in both high-energy ($E>$ 5 MeV) and low-energy ($E<$ 5 MeV) regions of open-shell nuclei $^{44,48,52,56}$Ca. In the high-energy region of these nuclei, proton and neutron ($1f_{7/2}$$\rightarrow$$1f_{5/2}$) transitions are the major contributors to the M1 strength at $T=1$ and 2 MeV. A finite contribution from $\pi$($1d_{5/2}$$\rightarrow$$1d_{3/2}$) transition is also observed at $T=2$ MeV in $^{40-60}$Ca nuclei. Also, new smaller peaks arise in the low-energy region ($E<$ 5 MeV) of neutron rich $^{48-60}$Ca isotopes as a result of neutron transitions, e.g.,  $\nu$($2p_{3/2}$$\rightarrow$$2p_{1/2}$) and $\nu$($1f_{7/2}$$\rightarrow$$1f_{5/2}$). 

Tables \ref{table1} and \ref{table2} display the proton and neutron isovector partial contributions ($b_{2qp}^{\pi(\nu)}$) to the M1 transition strength, as defined in Eq. (\ref{Part.Contri}), for the $^{52}$Ca nucleus. Table \ref{table1} pertains to the high-energy region, whereas Table \ref{table2} is provided for the low-energy region. It is evident from the tables that the number of excitation energies and related contributing transitions in the M1 response increases as one moves from $T = 0$ to 2 MeV, signifying the thermal unblocking of transitions. The reduced transition probability for a specific excited state, denoted as $B(M1,E)$, is determined by summing up the contributions of each configuration while considering their relative signs. First, we observe that temperature leads to the fragmentation of the excited states, impacts the contribution of the main configurations, and causes a decrease in the total $B(M1,E)$ value for a particular state with increasing temperature. Second, we found that new states appear; however, the contributions of these new configurations are either too small, or they interfere destructively, which, in turn, results in low strength. While temperature leads to the scattering of nucleons into the continuum, this effect is less pronounced for protons due to the Coulomb barrier. Hence, neutron transitions dominate in the low-energy region of M1 strength for neutron-rich Ca nuclei.

For further illustration, Figure \ref{fig3} displays the temperature dependence of the major high-energy $E_{\text{peak}}^{\text{High}}$ and low-energy $E_{\text{peak}}^{\text{Low}}$  peaks in the M1 strength distribution for $^{52}$Ca. In this case, $\nu$($1f_{7/2}$, $1f_{5/2}$) and $\nu$($2p_{3/2}$, $2p_{1/2}$) transitions are the primary contributors to the high-energy and low-energy peaks, respectively. Therefore, in Fig. \ref{fig3}, the SO splitting energies $\Delta \varepsilon_{LS}$ of these two respective configurations are shown for comparison relative to the M1 excitation energies. For the $E_{\text{peak}}^{\text{High}}$, we observe a decrease around the critical temperature $T_\textrm{c}$ = 0.528 MeV, which then continues to decrease gradually with increasing temperature. The SO splitting energy also decreases slightly. Although the impact of pairing is subtle in this region, its effect can be seen in the comparably rapid decrease in energy around the critical temperature. At higher temperatures, the slight decrease in the $E_{\text{peak}}^{\text{High}}$ is mainly related to the decrease in the SO splitting energy and weakening of the repulsive residual interaction. On the other hand, we observe a rapid decrease in the $E_{\text{peak}}^{\text{Low}}$ near the critical temperature, which is similar to the pairing phase transition, demonstrating the subtle interplay between the temperature effects and pairing correlations. Although the SO splitting energies are obtained as almost constant up to $T_\textrm{c}$, the decrease in the pairing correlations with increasing temperature leads to a sharp reduction of the energy. In other words, the pairing plays a significant role in the low-energy region of the M1 response of open-shell nuclei below $T_\textrm{c}$. At higher temperatures, $E_{\text{peak}}^{\text{Low}}$ continues to gradually decrease, but at a faster rate compared to $E_{\text{peak}}^{\text{High}}$ since the SO splitting energy of the relevant transition decreases more rapidly.

Next, we study the non-energy-weighted $m_0$ and energy-weighted $m_1$ moments of the M1 transition strength, calculated using the FT-RQRPA method. In Figure \ref{fig4}, we can see the behavior of $m_0$ (a) and $m_1$ (b) as a function of the mass number $A$ for Ca isotopes at temperatures ranging from $T=0$ to $2$ MeV. The M1 response does not appear when the SO-partner states are either fully occupied or completely empty, a characteristic stemming from the use of the one-body operator in Equation (\ref{M1 operator}). Consequently, the M1 response is absent at $T$=0 MeV in $^{40,60}$Ca and appears predominantly in other Ca isotopes through neutron transitions, e.g., $\nu$($1f_{7/2}\rightarrow1f_{5/2}$). It can be seen from Fig. \ref{fig4}(a) and \ref{fig4}(b) that the moments $m_0$ and $m_1$ have the highest magnitude for $^{48}$Ca and $^{52}$Ca, since the $\nu(1f_{7/2})$ (and $\nu(2p_{3/2})$ for $^{52}$Ca) state is nearly filled, and the corresponding M1 excitation strength is higher. For heavier isotopes of $^{56,60}$Ca, as the $\nu$($1f_{5/2}$) state begins to fill, the transition strength decreases due to the blocking of the $\nu$($1f_{7/2}\rightarrow1f_{5/2}$) transition, as can be seen in Figs. \ref{fig2} and \ref{fig4} at zero temperature. Similar results are also obtained in Ref. \cite{Oishi_20}.

With the increase in temperature, the $m_0$ and $m_1$ moments of $^{40,60}$Ca nuclei show a rapid increase above $T=$ 0.8 MeV due to the opening of new proton and neutron excitation channels. Although it is less pronounced compared to $^{40}$Ca and $^{60}$Ca, we also obtain an increase in the moments of $^{44}$Ca and $^{56}$Ca at higher temperatures. These results are also consistent with the results obtained in observation of Fig. \ref{fig2} and illustrate substantial contributions of new transitions which become allowed at higher temperatures. On the other hand, we obtain a moderate decrease in the moments for $^{48}$Ca and $^{52}$Ca at high temperatures. The main reason for the decrease in the $m_0$ and $m_1$ moments in $^{48}$Ca and $^{52}$Ca is the impact of the temperature on the $\nu$($1f_{7/2}$) and $\nu$($2p_{3/2}$) states. As can be seen from Tables \ref{table1} and \ref{table2}, the contribution of the $\nu$($1f_{7/2}$, $1f_{5/2}$) and $\nu$($2p_{3/2}$, $2p_{1/2}$) transitions to the M1 strength decreases substantially with increasing temperature, and the contribution of the new states is found to be subtle since their partial magnitudes are small and interfere destructively, resulting in a decrease in the M1 strength.

%%%%%%%%%%%%%%%%%%%%%%%%%%%%%%%%%%%%%%%%%%%%%%%%%%%%%%%%%%%%%%%%%
\begin{figure}[!ht]
\includegraphics[width=\linewidth,clip=true]{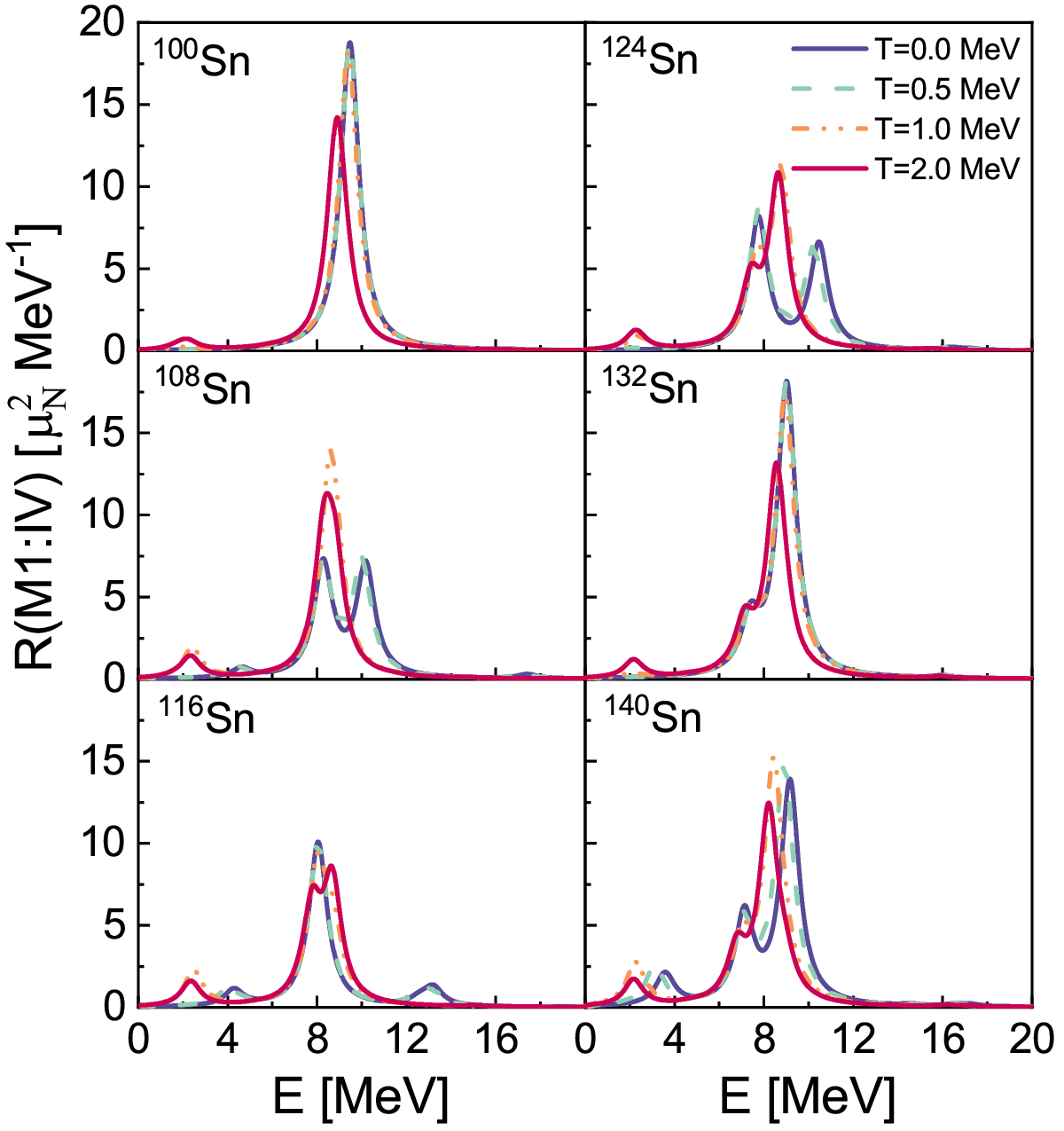}
\vspace{-0.5cm}
  \caption{The same as in Fig. \ref{fig2}, but for $^{100-140}$Sn isotopes.}
  \label{fig5}
\end{figure}
%%%%%%%%%%%%%%%%%%%%%%%%%%%%%%%%%%%%%%%%%%%%%%%%%%%%%%%%%%%%%%%%

%%%%%%%%%%%%%%%%%%%%%%%%%%%Table  M1 140Sn %%%%%%%%%%%%%%%%%%%
\begin{table*}[ht] \renewcommand{\arraystretch}{1.2}
\tabcolsep 0.15cm
\caption{Same as table \ref{table1} but for the high-energy M1 excitations in $^{140}$Sn nucleus. The excitation energies ($E$) are given in units of MeV.} 
\centering 
\begin{tabular}{l c c c c c c c c c c c c} 
\hline \\[-1.0em]
%   \multicolumn{10}{c}{High-lying region}  \\  \hline
      Configuration                 &     \multicolumn{2}{c}{$T=0$ MeV}     && \multicolumn{3}{c}{$T=1$ MeV}     && \multicolumn{5}{c}{$T=2$ MeV}  \\  % inserts table
                                               \cline{2-3}                         \cline{5-7}                          \cline{9-13} 
                                          & $E$=7.12  & $E$=9.16 && $E$=6.97  & $E$=8.41   & $E$=8.46 && $E$=6.79  & $E$= 8.04  &$E$=8.20 & $E$=8.27 & $E$=8.98 \\ 
\hline \\[-1.0em] 
$\nu$(1h$_{11/2}$$\rightarrow$1h$_{9/2}$) & \bf{-1.049} &\bf{3.664} && \bf{1.136}&\bf{2.382} &\bf{-2.743}&&\bf{1.041} &\bf{-1.857}&\bf{2.301}&\bf{-1.963}& \bf{0.437} \\
$\nu$(2f$_{7/2}$$\rightarrow$2f$_{5/2}$)  &  0.062      & 0.046     && -0.019    & 0.013     & -0.013    && -0.008    &  -0.003   & 0.006    & -0.005    & 0.003\\
$\nu$(2f$_{7/2}$$\rightarrow$3f$_{5/2}$)  & -0.020      & 0.037     && 0.031     & 0.016     & -0.016    && 0.014     &  -0.007   & 0.012    & -0.009    & 0.003\\
$\nu$(1h$_{9/2}$$\rightarrow$2h$_{11/2}$) &  0.001      &           && -0.003    & -0.028    &           &&           &  -0.013   &\bf{0.111}&\bf{0.145} & -0.003\\
$\nu$(1i$_{13/2}$$\rightarrow$i$_{11/2}$)&              &           && -0.008    & 0.004     & -0.003    && -0.049    &  -0.008   & 0.015    & -0.013    & 0.005\\
$\nu$(1i$_{13/2}$$\rightarrow$2i$_{11/2}$)& -0.001      &           && 0.005     & -0.029    & 0.031     && 0.022     &  0.038    &-0.087    & 0.089     & \bf{0.890}\\
$\pi$(1g$_{9/2}$$\rightarrow$1g$_{7/2}$)  & \bf{3.913}  &\bf{0.828} &&\bf{-3.227}&\bf{0.748} &\bf{-0.806}&&\bf{-3.183}&\bf{-0.281}&\bf{0.505}&\bf{-0.477}& \bf{0.158} \\
$\pi$(1h$_{11/2}$$\rightarrow$1h$_{9/2}$) &             &           &&           & 0.006     &           && 0.029     &\bf{0.609} &\bf{0.298}&\bf{-0.179}& 0.019 \\ \hline
total $B(M1,E)$ [$\mu_{N}^{2}$]           & 8.448       & 20.585    && 4.420     & 9.627     & 12.403    && 4.647     & 2.313     & 9.979    & 5.811     & 2.281 \\
                  
\hline \\ [-1.ex]
\end{tabular}
\label{table3} 
\end{table*}
%%%%%%%%%%%%%%%%%%%%%%%%%%%%%%%%%%%%%%%%%%%%%%%%%%%%%%%%%%

\subsection{Tin isotopes}
The temperature dependence of isovector M1 strength distributions is also systematically studied in even-even $^{100-140}$Sn isotopes, as shown in Fig. \ref{fig5}. First, we consider the $T=0$ MeV case. The M1 excitations are expected due to the proton $\pi(1g_{9/2}\rightarrow1g_{7/2})$ and the neutron $\nu(1g_{9/2}\rightarrow1g_{7/2})$, $\nu(2d_{5/2}\rightarrow2d_{3/2})$, and/or $\nu(1h_{11/2}\rightarrow1h_{9/2})$ transitions until the higher state in the SO configuration becomes fully occupied to block the M1 transitions. In closed shell nuclei
$^{100}$Sn and $^{132}$Sn, the pairing correlations do not contribute, hence, the M1-excitation energy is determined mainly by the SO splitting energy as well as the residual interaction. For $^{100}$Sn at zero temperature, the M1 strength is obtained as a single peak at 9.47 MeV, which is mainly formed with the proton and neutron ($1g_{9/2}$ and $1g_{7/2}$) transitions. As the number of neutrons increases along the Sn isotope chain, the M1 response exhibits two peaks in the high-energy region. The first peak at lower energy is primarily due to proton transitions, while the second, at higher energy, is dominated by neutron transitions. Similar results are obtained for the $T=$ 0.5 MeV case, where M1 strength distribution shifts slightly to the lower energies, as shown in Fig. \ref{fig5}. A thorough study of M1 transitions in $^{100-136}$Sn isotopes has been carried out at $T$ = 0 MeV using the RQRPA framework, and a detailed description is given in Ref. \cite{PhysRevC.103.054306}.
%%%%%%%%%%%%%%%%%%%%%%%%%%%%%%%%%%%%%%%%%%%%%%%%%%%%%%%%%
\begin{table}[ht!] \renewcommand{\arraystretch}{1.2}
\tabcolsep 0.04cm
\caption{Same as Table \ref{table1} but for the low-energy M1 excitations in $^{140}$Sn nucleus.} 
\centering 
\begin{tabular}{l c c c}  \hline \\[-1.0em]
 %  \multicolumn{5}{c}{Low-lying region}  \\ \hline
   Configuration                          & $T=0$ MeV     & $T=1$ MeV      & $T=2$ MeV  \\
                                                                          
                                          & $E$=3.57 MeV  & $E$=2.26 MeV    & $E$=2.17 MeV  \\ \hline
$\nu$(2f$_{7/2}$$\rightarrow$2f$_{5/2}$)  & \bf{2.078}    & \bf{-2.267}     & \bf{-1.434}    \\
$\nu$(1h$_{11/2}$$\rightarrow$1h$_{9/2}$) & -0.138        & 0.102           &  0.079      \\
$\nu$(2d$_{5/2}$$\rightarrow$2d$_{3/2}$)  &               &                 & \bf{-0.166}    \\
$\pi$(2d$_{5/2}$$\rightarrow$2d$_{3/2}$)  &               & -0.016          & -0.131   \\ 
$\pi$(1g$_{9/2}$$\rightarrow$1g$_{7/2}$)  & \bf{-0.267}   & \bf{0.155}      & 0.092     \\\hline
total $B(M1,E)$[$\mu_{N}^{2}$]            &  2.815        &  4.088          & 2.403     \\
\hline \\ [-1.ex]
\end{tabular}
\label{table4} 
\end{table}
%%%%%%%%%%%%%%%%%%%%%%%%%%%%%%%%%%%%%%%%%%%%%%%%%%%%%%%%%%%%%%%%%%%%%%%%%%%%%%%

Similar to the findings in Ca isotopes, the M1 strength starts to shift downward, and new excited states appear both in the low-energy and high-energy regions with increasing temperature. Besides the downward shift in the M1 strength, we also observe a change in the $^{108,124,140}$Sn nuclei from a two-peak to a single-peak structure of M1 strength as the temperature increases. Due to the vanishing of pairing correlations at critical temperatures, the configuration energies between the proton and neutron components start to approach one another, which results in the merging of two peaks into a single peak at higher temperatures. To elucidate further, the structure of M1 strength distributions at finite temperatures is explored by analysing all transitions that contribute to these states for each considered Sn nucleus at $T=1$ and 2 MeV. This study has revealed that along with the major $\nu, \pi$$(1g_{9/2}\rightarrow1g_{7/2})$ transitions, $\nu,\pi$$(1h_{11/2}$$\rightarrow$$1h_{9/2}$) transitions also start contributing in the high-energy region of $^{100-140}$Sn nuclei at higher temperatures. A finite contribution from $\nu$($1i_{13/2}$$\rightarrow$$2i_{11/2}$) and $\nu$($1h_{9/2}$$\rightarrow$$2h_{11/2}$) transitions are also obtained in the neutron-rich Sn isotopes due to the thermal unblocking. Moreover, a new low-energy M1 peak below $E=$ 5 MeV appears in all Sn isotopes at finite temperatures. In the lower-mass Sn isotopes, the $\nu$($2d_{5/2}$$\rightarrow$$2d_{3/2}$) transition is the major contributor to the low-energy peak, and $\nu$($2f_{7/2}$$\rightarrow$$2f_{5/2}$) transition also begins contributing in the neutron-rich nuclei. A thorough analysis of the partial contributions of protons and neutrons in the high-energy ($E> 5$ MeV) and low-energy  regions ($E< 5$ MeV) is given in Tables \ref{table3} and \ref{table4}, respectively, for the neutron-rich $^{140}$Sn nucleus at $T=0$, 1, and 2 MeV.

%%%%%%%%%%%%%%%%%%%%%%%%%%%%%%%%%%%%%%%%%%%%%%%%%%%%%%%%%%%%%%%%%%%%%%%%%%%%
\begin{figure*}[ht!]
\begin{center}
\includegraphics[width=\linewidth,clip=true]{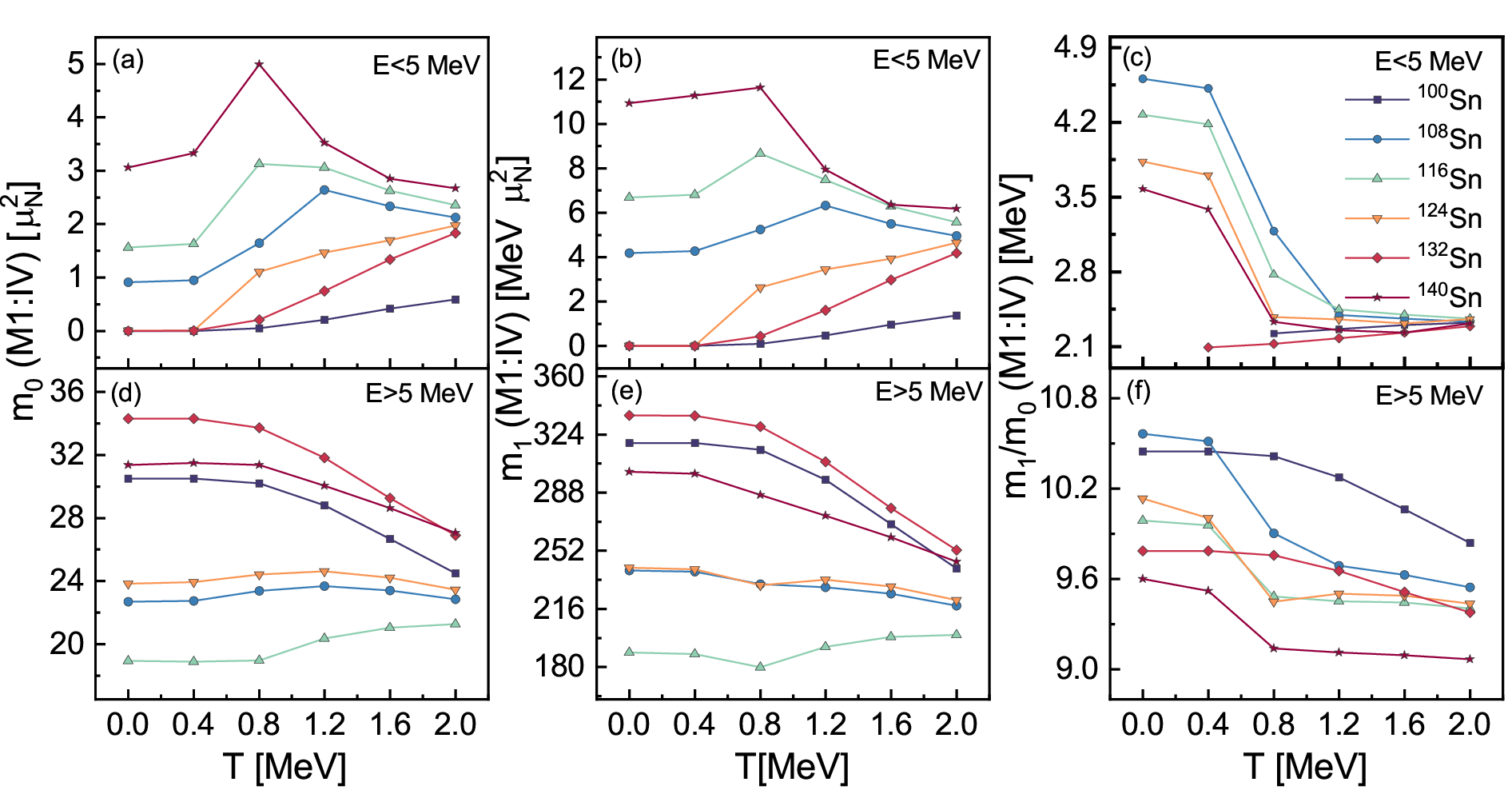}
  \end{center} \vspace{-0.7cm}
  \caption{The isovector M1 non-energy-weighted sum $m_0$, energy-weighted sum $m_1$ and the centroid energy ($m_1/m_0$) in $^{100-140}$Sn for the low-energy region $E<$ 5 MeV [panels (a)-(c)] and high-energy region $E>$ 5 MeV [panels (d)-(f)] as a function of temperature.}
  \label{fig6}
\end{figure*}
%%%%%%%%%%%%%%%%%%%%%%%%%%%%%%%%%%%%%%%%%%%%%%%%%%%%%%%%%%%%%%%%%%%%%

This investigation extends further to the M1 non-energy-weighted $m_0$ and energy-weighted $m_1$ summations, along with the centroid energies ($E_c = m_1/m_0$), as functions of temperature for $^{100-140}$Sn isotopes, as shown in Fig. \ref{fig6}. The upper panels (a-c) display moments for the low-energy region ($E < 5$ MeV), while the lower panels (d-f) represent these summations for the high-energy region ($E > 5$ MeV). This energy cut-off is selected because it clearly separates the low- and high-energy regions in the M1 strength distributions for all Sn isotopes, as shown in Fig. \ref{fig5}.
We then elaborate on the variation of moments in the low-energy region with increasing temperature, as presented in Figs. \ref{fig6}(a-c). For closed-shell $^{100,132}$Sn nuclei, the moments and centroid energy remain nearly zero up to $T = 0.8$ MeV because the spin-orbit partner states are unavailable for M1 transitions in the low-energy region. Then, $E_c$ values increase linearly with temperature due to the thermal unblocking of these states. The rate of increase in the moments is higher for $^{132}$Sn compared to $^{100}$Sn because it is neutron-rich, and with increasing temperature particles are more easily scattered to the continuum region. Additionally, we observed a slight increase in the centroid energies with increasing temperature.
However, the behavior of the $m_0$ and $m_1$ moments in open-shell nuclei with increasing temperature is not straightforward and displays dependence on the particular shell structure. In all open-shell nuclei, a sharp increase is observed in the moments above $T=$ 0.4 MeV and up to the critical temperatures. This change is mainly related to the decrease in the pairing correlations. At higher temperatures, pairing correlations disappear and a decrease is observed in the $m_0$ and $m_1$ moments for $^{108,116,140}$Sn. Only for $^{124}$Sn, the moments continue to increase with increasing temperature. For instance, in $^{140}$Sn, we found that the impact of the major configuration $\nu$(2f$_{7/2}$$\rightarrow$2f$_{5/2})$ decreases with increasing temperature due to the weakening of the residual interaction, and the partial contributions from the new configurations interfere destructively. As a result, we observe a decrease in the moments, as illustrated in Table \ref{table4}. Similar results are also obtained for $^{108,116,140}$Sn. For $^{124}$Sn, there is a negligible M1 response below $E<5$ MeV at very low temperatures due to the inaccessibility of single-particle (SP) partners for M1 transitions in the low-energy region. However, due to thermal unblocking, new excitation channels primarily emerge between the $\nu$($2d_{5/2}$, $2d_{3/2}$) configurations, and the sum of moments $m_0$ and $m_1$ also increases with temperature. It is also observed that the difference between the $m_0$ and $m_1$ moments of different nuclei decrease as the shell effects diminish at higher temperatures. Finally, the centroid energy first decreases sharply up to the critical temperatures and then exhibits only slight changes as the temperature increases further.

In the high-energy region ($E >$ 5 MeV), the isovector variables $m_0$, $m_1$, and $E_c$ smoothly reduce for closed-shell nuclei $^{100,132}$Sn. This means that at higher temperatures, a portion of the M1 strength shifts to the low-energy region. However, a transition in the variation of moments and centroid energies is observed around $T \approx$ 0.8 MeV (or critical temperatures) for open-shell Sn isotopes in the high-energy region. At higher temperatures, we observed either a slight increase or decrease in the $m_0$ and $m_1$ moments, depending on the particular nuclei. For all open-shell nuclei, the centroid energy decreases with increasing temperature up to the critical point, after which it changes only slightly with further temperature increases. As discussed earlier in the explanation of Fig. \ref{fig5}, this transition occurs due to the interplay of pairing and thermal effects. The behavior of the M1 strength is governed by subtle effects related to single-particle structure, pairing correlations, respective occupation probabilities, and the FT-RQRPA residual interaction. 

%%%%%%%%%%%%%%%%%%%%%%%%%%%%%%%%%%%%%%%%%%%%%%%%%%%%%%%%%%%%%%%%%%%%%
\section{Summary} \label{Summary}
A novel approach, the fully self-consistent FT-RQRPA framework based on the relativistic energy density functional, has been applied to study the temperature dependence of isovector M1 excitations. The properties of $^{40-60}$Ca and $^{100-140}$Sn isotopes were calculated at finite temperatures within the relativistic framework using the FT-HBCS with the DD-PCX functional and separable pairing interaction. The FT-RQRPA is extended for the study of unnatural parity excitations of M1 type.

By employing the FT-HBCS, we studied the evolution of spin-orbit (SO) splitting energies in the open-shell nucleus $^{52}$Ca at temperatures both below and above the critical temperature $T_\textrm{c}$. We observed a reduction in the spin-orbit splitting energies as the temperature increased. The FT-RQRPA calculation of M1 response in both Ca and Sn isotopes showed a systematic shift of the strength distributions to lower energies in both Ca and Sn isotope chains as the temperature increased. The changes in the M1 response are influenced not only by the weakening and disappearance of the pairing correlations and the softening of the repulsive residual interaction but also by the decreasing SO splitting energies. Remarkably, we found that M1 strength suddenly emerges as the temperature increases in nuclei with initially blocked M1 transitions. This phenomenon is attributed to the thermal unblocking of new proton and neutron excitations between the spin-orbit partner states. The thermal unblocking of states also results in the formation of new M1 peaks in the low-energy region.

The results of our study are also quantitatively assessed by analyzing the temperature dependence of the non-energy-weighted sum $(m_0)$, the energy-weighted sum of the transition strength $(m_1)$, and the centroid energy in the considered isotope chains. The changes in the moments $m_0$ and $m_1$ of Ca and Sn nuclei with increasing temperature indicate that the isotopic evolution of the M1 response lacks a uniform pattern, at least at low temperatures. This lack of uniformity can be attributed to the influence of multiple contributing factors, such as the particular shell structure, pairing correlations, and residual interaction. In general, we observe a sharp change in the variation of moments and centroid energies with temperature for open-shell Sn nuclei. This change results from a phase transition from a superfluid state to the normal state that occurs near the critical temperatures, where pairing effects decrease sharply. At high temperatures, the shell effects also diminish, and the differences between the $m_0$ and $m_1$ moments decrease for the considered isotope chains. In conclusion, we have found that temperature effects can considerably modify the magnetic dipole response. Further studies on the possible contributions of magnetic transitions at finite temperature in the $\gamma$-strength functions for $(n,\gamma)$ reactions are required. These studies are left for future work.

%%%%%%%%%%%%%%%%%%%%%%%%%%%%%%%%%%%%%%%%%%%%%%%%%%%%%%%%%%%%%%%%%%%%%
\section{Acknowledgements} 
This work is supported by the QuantiXLie Centre of Excellence, a project co-financed by the Croatian Government and European Union through the European Regional Development Fund, the Competitiveness and Cohesion Operational Programme (KK.01.1.1.01.0004). E.Y. acknowledges support from the Science and
Technology Facilities Council (UK) through grant ST/Y000013/1.

%%%%%%%%%%%%%%%%%%%%%%%%%%%%%%%%%%%%%%%%%%%%%%%%%%%%%%%%%%%%%%%%%%%%%
\bibliographystyle{apsrev4-2}
\bibliography{M1.bib}
%%%%%%%%%%%%%%%%%%%%%%%%%%%%%%%%%%%%%%%%%%%%%%%%%%%%%%%%%%%%%%%%%%%%%%

\end{document}